\documentclass[twocolumn]{aastex61}
\pdfoutput=1 %for arXiv submission
\usepackage{amsmath,amstext}
\usepackage[T1]{fontenc}
\usepackage{apjfonts} 
\usepackage[figure,figure*]{hypcap}

\shorttitle{Unified TDE model}
\shortauthors{Dai et al.}

\begin{document}

\title{A unified model for tidal disruption events}

\correspondingauthor{Lixin Dai}
\email{lixin.dai@nbi.ku.dk}

\author{Lixin Dai}
\affiliation{DARK Cosmology Centre, Niels Bohr Institute, University of Copenhagen, Juliane Maries Vej 30, 2100 Copenhagen {\O}, Denmark}
\affiliation{Department of Physics and Joint Space-Science Institute, University of Maryland, College Park, MD 20742, USA}
\author{Jonathan C. McKinney}
\affiliation{Department of Physics and Joint Space-Science Institute, University of Maryland, College Park, MD 20742, USA}
\author{Nathaniel Roth}
\affiliation{Department of Astronomy and Joint Space-Science Institute, University of Maryland, College Park, MD 20742, USA}
\author{Enrico Ramirez-Ruiz}
\affiliation{Department of Astronomy and Astrophysics, University of California Santa Cruz, 1156 High Street, Santa Cruz, CA 95060, USA}
\affiliation{DARK Cosmology Centre, Niels Bohr Institute, University of Copenhagen, Juliane Maries Vej 30, 2100 Copenhagen {\O}, Denmark}
\author{M. Coleman Miller}
\affiliation{Department of Astronomy and Joint Space-Science Institute, University of Maryland, College Park, MD 20742, USA}

\begin{abstract}

In the past few years wide-field optical and UV transient surveys as well as X-ray telescopes have allowed us to identify a few dozen candidate tidal disruption events (TDEs). While in theory the physical processes in TDEs are expected to be ubiquitous, a few distinct classes of TDEs have been observed. Some TDEs radiate mainly in NUV/optical while others produce prominent X-rays. Moreover, relativistic jets have been observed in only a handful of TDEs. This diversity might be related to the details of the super-Eddington accretion and emission physics relevant to TDE disks. In this Letter, we utilize novel three-dimensional general relativistic radiation magnetohydrodynamics simulations to study the super-Eddington compact disk phase expected in TDEs. Consistent with previous studies, geometrically thick disks, wide-angle optically-thick fast outflows and relativistic jets are produced. The outflow density and velocity depend sensitively on the inclination angle, and hence so does the reprocessing of emission produced from the inner disk. We then use Monte-Carlo radiative transfer to calculate the reprocessed spectra and find that that the observed ratio of optical to X-ray fluxes increases with increasing inclination angle. This naturally  leads to  a unified model for different classes of TDEs in which the  spectral properties of the TDE depend mainly on the viewing-angle of the observer with respect to the orientation of the disk.

\end{abstract}


\section{Introduction} \label{sec:intro}

Stars in a galaxy can occasionally be scattered to approach the central supermassive black hole (SMBH) so close that they can be tidally disrupted. About half of the stellar mass can be accreted onto the black hole, producing a luminous flare. Theoretical foundations for such tidal disruption events (TDEs) have been laid out since the 1970s \citep[e.g.,][]{Hills75, Rees88}. The rate that debris falls back to the black hole is expected to increase in $\sim1$ month, and then declines with time as $t^{-5/3}$ in $\sim1$ year \citep{Phinney89, Evans89, Lodato09, Guillochon13}. Furthermore, if the luminosity is Eddington limited and the size of the photosphere is similar to the tidal disruption radius, then the effective temperature of the thermal emission is $\sim~{\rm few}~\times~10^5$ K, which means the flare should be bright in soft X-ray \citep{Cannizzo90, Ulmer99}. 

A few TDEs were identified in the 1990s using the soft X-ray satellite ROSAT \citep{Komossa15review}. Then recently a few dozen TDEs were found through mainly optical and UV wide-field transient surveys. The recent breakthrough in TDE observations, however, demonstrated that our theoretical understanding is incomplete. First, many newly discovered TDEs produced luminous optical/NUV flares with a temperature of $\sim10^4$K \citep[e.g.,][]{Gezari12, Arcavi14}. Some of them produced prominent X-ray emission corresponding to $\sim10^5$K besides optical emission \citep{Miller15, Holoien16_14li, Holoien16_15oi}. Second, only three TDEs have been observed to produce transient relativistic jets \citep[e.g.,][]{Bloom11,Burrows11, Levan11,Cenko12, Brown15}. Since the disruption and accretion physics in TDEs are expected to be quasi-universal, it is intriguing to ask why similar disruption conditions can give rise to distinct types of TDEs. 

The first task in constructing a general scheme for interpreting TDEs is to decide which parameters exert a controlling influence upon their observed properties. 
The mass of the SMBH dictates a characteristic duration and luminosity scale for TDE activity, though probably not the qualitative character. The spin parameter $a\equiv cJ/GM_{\rm BH}^2$ ($J$ and $M_{\rm BH}$ are respectively the angular momentum and mass of the hole, $c$ is the speed of light, and $G$ is the gravitational constant) may control whether a jet can be produced and be important for interpreting the observed properties of jetted TDEs \citep{Tchekhovskoy14}. In the present study, we nonetheless argue that the orientation relative to our line of sight may turn out to be the key parameter for interpreting the properties of most observed TDEs.

To examine this, we study an epoch close to peak of the flare. For most TDEs the peak of the fallback rate is 1-2 orders of magnitude above the Eddington accretion rate defined as $\dot{M}_{\rm Edd} = L_{\rm Edd}/(\eta_{\rm NT}~c^2)$, where $L_{\rm edd}=4\pi GM_{\rm BH}c/\kappa$ is the Eddington luminosity for an opacity $\kappa$, and $\eta_{\rm NT}$ is the nominal accretion efficiency for the Novikov-Thorne thin disk solution \citep{Novikov73} ($\eta_{\rm NT} = 12.2\%$ for $a=0.8$ used in the simulation for this study).
If the debris can quickly reduce orbital energy and assemble a disk, then its accretion rate onto the black hole can also be super-Eddington. Hydrodynamical simulations \citep{Shiokawa15,Bonnerot16, Hayasaki16} and semi-analytical studies \citep{Dai13,Dai15,Guillochon15} have shown that only in certain parameter regions can fast disk assembly happen. However, the light curves of most observed TDEs have exhibited temporal pattern similar to the fallback rate, indicating short circularization and viscous timescales \citep{Mockler18}. Therefore, in the observed events, debris is likely supplied to the SMBH in a super-Eddington fashion.

Here we present the results of the first realistic simulation to understand the super-Eddington accretion and emission physics in TDEs. It has been predicted that in super-Eddington accretion photons are trapped within the accretion flow and a geometrically thick accretion disk forms due to large radiation pressure \citep{Begelman78, Abramowicz88}. Recently the development of novel radiation magnetohydrodynamics (MHD) codes, some of which also performed under full general relativity (GR), have helped us understand more about such accretion flow \citep[e.g.,][]{Ohsuga05, Jiang14, McKinney14, Sadowski15}. These works have demonstrated that wide-angle fast outflows are launched from the disk. Also, if large-scale ordered magnetic fluxes are provided to the accretion flow around a spinning black hole, a relativistic jet can be produced magnetically \citep{McKinney15}. 

\begin{figure*}[ht!]
\figurenum{1}
\label{fig:basic}
\plotone{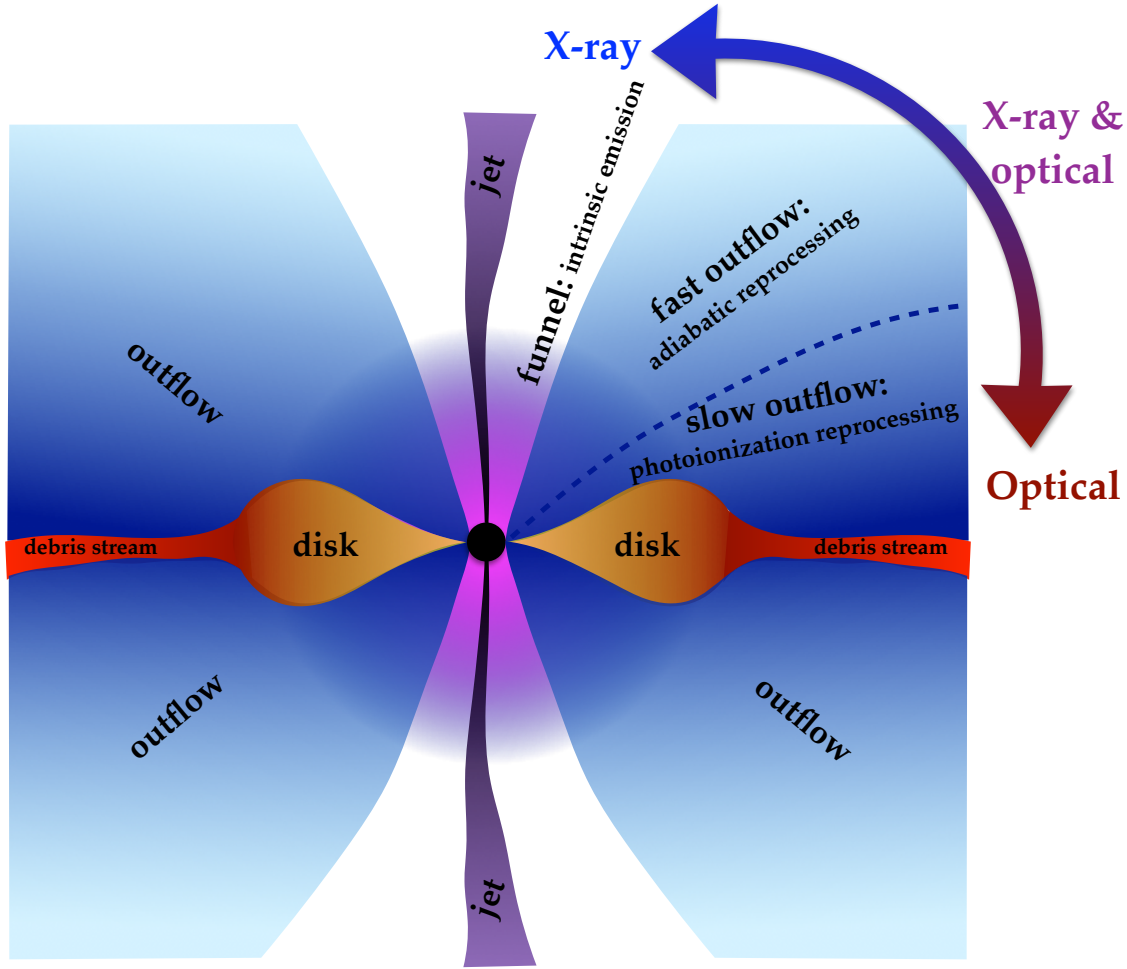}
\caption{A schematic picture showing the viewing-angle dependence for the observed emission from a TDE super-Eddington disk. The emission from the inner disk is reprocessed by the optically thick outflows and outer disk. Only when the observer is looking into the optically thin funnel is the inner disk exposed, which can reveal strong, beamed X-ray and EUV radiation. Otherwise, X-ray is reprocessed into optical/NUV emission via photoionization (in denser outflow or disk at high inclination angles) or adiabatic cooling (in ultrafast outflow at low inclination angles).  A jet is included in the picture for completeness, though most TDEs may not produce jets.}
\end{figure*}

Previous super-Eddington simulations have mostly focused on extended disks around stellar-mass black holes. In order to study TDE disks, we simulate a compact super-Eddington accretion disk around a SMBH, using a 3D fully general relativistic radiation magnetohydrodynamics (GRRMHD) code (section \ref{sec:GRRMHDsetup}). We then post-process the simulation data for radiative transfer analysis using a Monte-Carlo code (section \ref{sec:RTsetup}). We illustrate the qualitative results in the schematic Fig.\ref{fig:basic} and give details in section \ref{sec:results}. Our summary, with caveats and a discussion of future work is found in section \ref{sec:summary}.

\section{TDE super-Eddington accretion: methodology}
\label{sec:simulation}

Reprocessing of emission by an optically thick envelope, such as the outflows from super-Eddington accretion, has been discussed in \citet{Loeb97, Strubbe09, Coughlin14, Metzger16, Roth16}. In particular, \citet{Metzger16} and \citet{Roth16} suggested that there could be a viewing angle dependence for emission, though they still used a spherically symmetric envelope with ad-hoc profile for calculations. A general relativistic simulation of a super-Eddington TDE disk is the key to provide a clear, qualitative understanding of the outflow profile and the viewing-angle dependence of the observed emission.

\subsection{Fully 3D GR Radiation MHD simulation set up}
\label{sec:GRRMHDsetup}
We simulate a super-Eddington TDE disk using the fully 3D general relativistic radiation magnetohydrodynamics (GRRMHD) code called {\tt HARMRAD} \citep{McKinney14}, which treats radiation under M1 closure \citep{Levermore84}. The gas is assumed to have solar chemical abundances (mass fractions of H, He, and ``metals'', respectively, $X = 0.7, Y = 0.28, Z = 0.02$), so its electron scattering opacity is $\kappa_{\rm es}~=~0.2(1+X)~{\rm cm^2~g^{-1}}$. For absorption and emission, frequency-mean opacities are used (see \citet{McKinney15} for the expression; also the Chianti opacity is turned off as it is unimportant for the TDE disk temperature). Thermal Comptonization is also included. There are 128 cells in radius $r$ from $1.2~R_g$ to $10^5R_g$ ($R_g=GM_{\rm BH}/c^2$ is the gravitational radius of the black hole) with cell size increasing exponentially until $r \approx 500~R_g$ and then even faster, 64 cells in $\theta$-grid from 0 to $\pi$ with finer resolution in the jet and disk region, and 32 cells in $\phi$-grid spanning uniformly from 0 to $2\pi$. We provide large-scale poloidal magnetic flux to the initial disk with plasma beta of $\sim 20-30$.

We tailor the parameters and initial conditions to be study TDE disks. The SMBH has a mass $M_{\rm BH}~\approx~5\times10^6~M_\odot$ and spin parameter $a=0.8$. The initial disk is Keplerian with a rest-mass density that is Gaussian in angle with a height-to-radius ratio of $H/R\approx0.3$. Radially the density follows a power-law of $\rho\propto~r^{-1.3}$ out to $R_{\rm disk}=500~R_g$ and exponentially decays with $r>R_{\rm disk}$. We set up a small disk, since in TDEs stellar debris is provided to the SMBH from a close distance, which makes TDE disks much more compact than AGN or X-ray binary disks. We note that the total mass and angular momentum of our simulated disk may not be the same as a TDE disk. The current disk profile has been selected from several test runs, so that the disk, while being as compact as possible, can sustain until inflow equilibrium is achieved to a large enough radius. However, as long as the accretion rate is consistent with a typical TDE accretion rate, then because the radiation, outflows and other energy output, which are mostly generated from very close the black hole, are governed by the accretion rate, they are therefore realistic. The compact size of the initial disk allows us to calculate how outflows warp around the disk and how the emission is reprocessed at high inclination angles.

\subsection{Radiative transfer setup}
\label{sec:RTsetup}
To determine the escaping radiation spectrum as a function of viewing angle, we perform Monte Carlo radiative transfer calculations using the code {\tt SEDONA} \citep{Kasen06}, modified to include a solution to the non-local thermodynamic equilibrium equations of ionization bound-electron level populations \citep{Roth16} and Comptonization \citep{Roth17}, although we neglect the effects of stimulated scattering. We track free-free absorption and emission, along with bound-free and bound-bound interactions with H, He, and O in solar abundance ratios. 

The radiative transfer tracks the photon propagation in 3D, but the gas densities and velocities are treated in spherical symmetry. We explore the inclination angle dependence of the problem by performing four separate calculations where the gas density and velocity correspond to an average of those quantities within a range of inclination angles from the GRRMHD simulations. For the velocity averaging, we consider contributions from outflows and disk outside the inner boundary. We label these bins 1 through 4: bin 1 corresponds to angles between $67.5^{\circ}$ and $87.4^{\circ}$ from the pole, bin 2 between $45^{\circ}$ and $67.5^{\circ}$, bin 3 between $22.5^{\circ}$ and $45^{\circ}$, and bin 4 between $5.7^{\circ}$ and $22.5^{\circ}$.

The gas temperatures are recomputed within the Monte Carlo calculation under the assumption of radiative equilibrium, and these along with the gas ionization state and bound electron states are solved iteratively along with the radiative transfer solution as in \citet{Roth16}. A fixed amount of radiative energy is injected from the inner boundary of the calculation at each time step, and any photon packets that are directed back within the inner boundary are removed from the calculation. We place the inner boundary at the radial location where the $\theta$-averaged velocity field transits from inflow to outflow - this is typically on the order of a few $R_g$. For all four bins we set the spectrum at the inner boundary to correspond to a blackbody at $T\sim 10^6$ K - this is approximately equal to the color temperature of the radiation at those radii as computed in the GRRMHD simulation.

\section{Results and Connections to Observations: dynamics, energy, emission}
\label{sec:results}

\subsection{GRRMHD Simulation Results}

\begin{figure*}
\figurenum{2}
\label{fig:zoomin}
\plotone{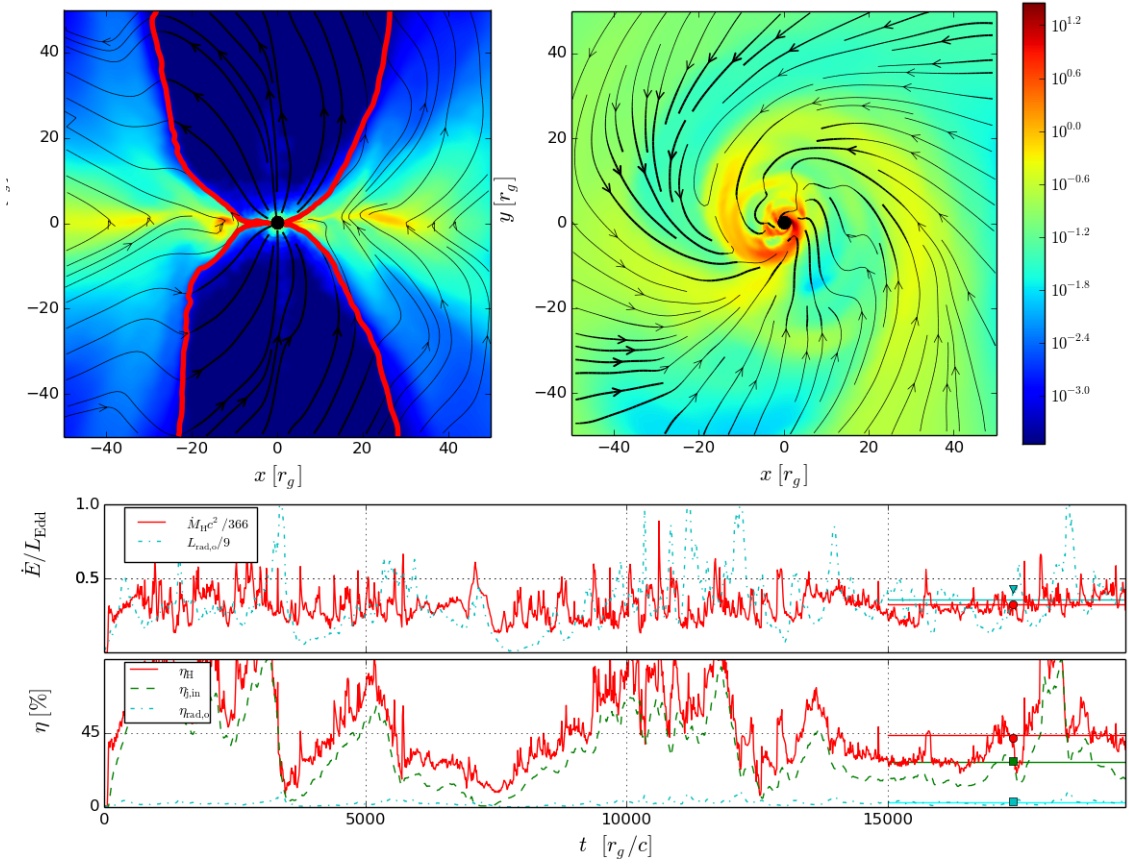}
\caption{Evolved snapshot (at $t = 17400~R_g/c$) showing $\log_{10}$ of rest-mass density (scaled by an Eddington density of $\rho_{\rm Edd}~=~\dot{M}_{\rm Edd}/(4~\pi/3~{r_g}^3)\sim~4\times~10^{-12}~{\rm g~cm^{-3}}$, shown in color with legend on right) in both the z-x plane (top-left panel) and y-x plane (top-right panel). Black lines trace field lines, where thicker black lines show where field is lightly mass-loaded. In the top-left panel, the thick red line corresponds to the jet boundary where the electromagnetic energy equals the rest-mass energy of the gas. The bottom panel has two sub-panels. Horizontal solid lines show the averages over the  period from $15000~R_g/c$ until the end of the simulation, while square/circle tickers are placed at the given time and values. The top sub-panel shows $\dot{M}$ through the black hole horizon and radiative luminosity ($L_{\rm rad,o}$, from optically thin region at $r\sim~1000~R_g$). All quantities have been normalized by the Eddington luminosity, where in addition the mass flux has been divided by $\approx366$ and radiative luminosity by $\approx9$, so that both quantities can be shown on a single panel. The bottom sub-panel shows the efficiencies, where $\eta_{H}$ is the total efficiency, $\eta_{\rm j, in}$ is the jet efficiency, and $\eta_{\rm rad,o}$ is the radiative efficiency.}
\end{figure*}

\begin{figure*}
\figurenum{3}
\label{fig:outflow}
\plottwo{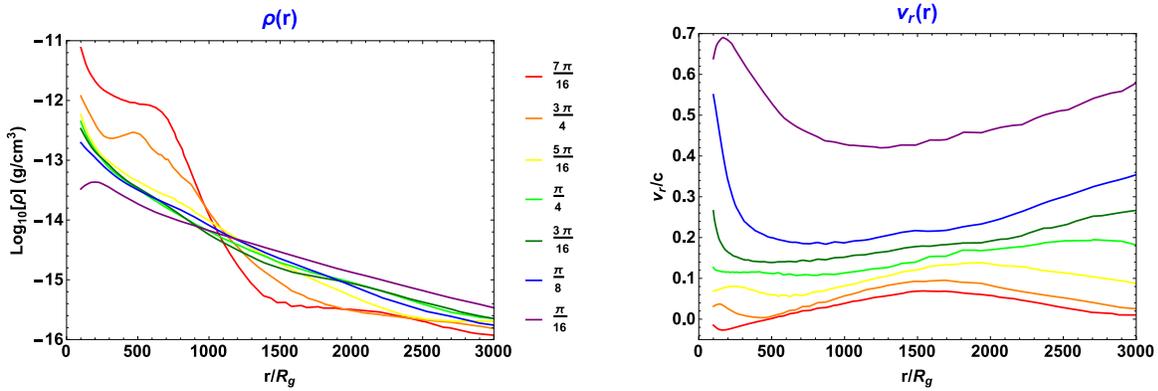}{VrR.pdf}
\caption{The $\phi$- and time-averaged density and radial velocity of the gas, as functions of $r$, along different inclination angles. Closer to the equatorial plane (red), the gas flow is denser close to the black hole and moves out slower. When looking down the funnel close to the pole (purple), the gas is much more dilute, and moves outward extremely fast (at a few$\times 0.1~c)$. 
}
\end{figure*}

\begin{figure*}
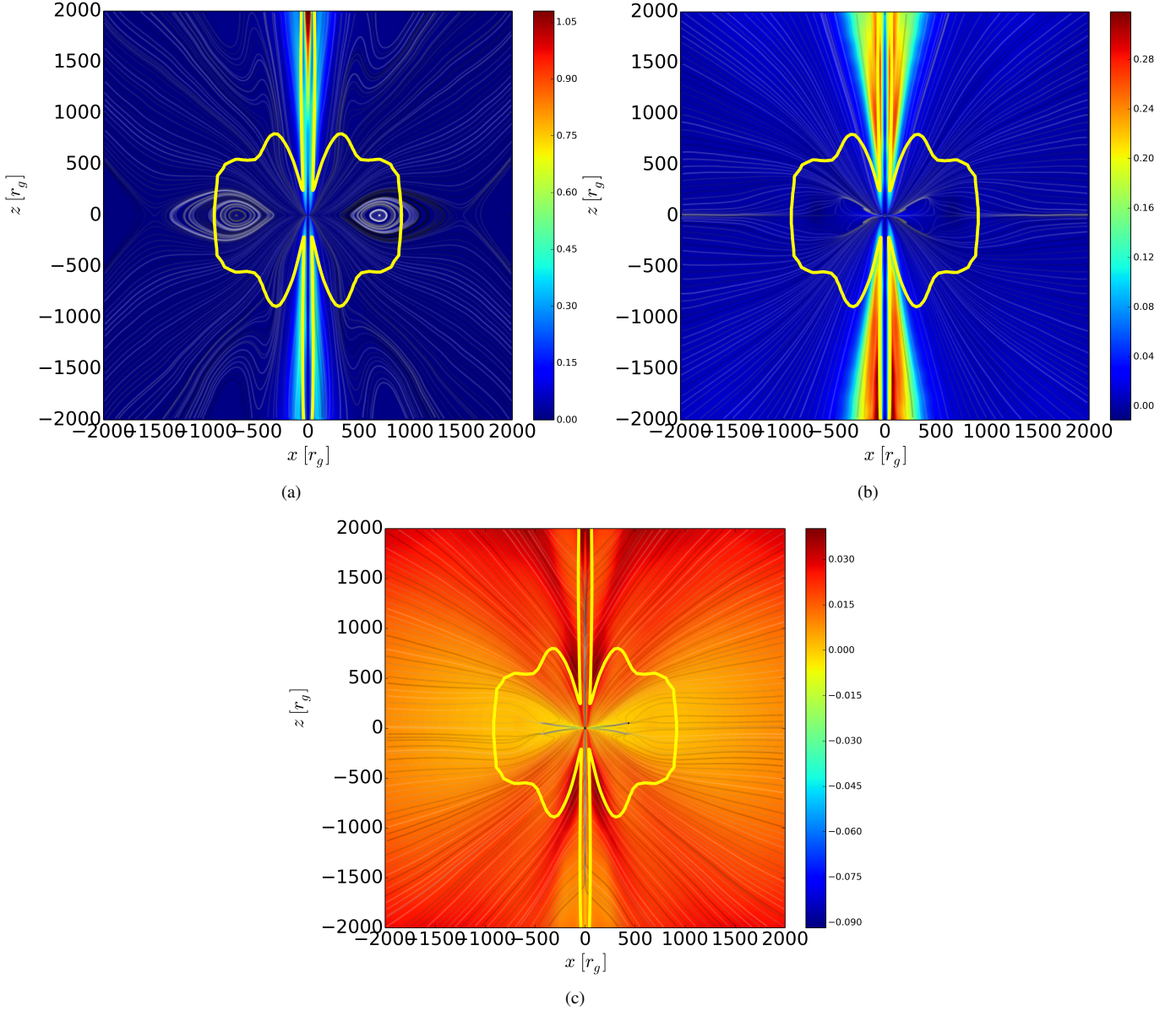

\figurenum{4}
\gridline{\fig{fig2a.png}{0.5\textwidth}{(a)}
		  \fig{fig2b.png}{0.5\textwidth}{(b)}
         }
         {\fig{fig2d.png}{0.5\textwidth}{(c)}
         }
\caption{The left panel (a) shows time-$\phi$-averaged magnetic flux lines (translucent gray lines) with electromagnetic luminosity per unit angle ($dL_{\rm EM}/d\theta)/(\dot{M_H}c^2)~\sim~(dL_{\rm EM}/d\theta)/(120~L_{\rm Edd}$) (color with legend), with blue contour showing the jet boundary, and yellow contour showing the electron scattering photosphere. The image is duplicated across the x = 0 line. The middle panel (b) shows the gas velocity flux lines and kinetic + gravitational energy luminosity per unit angle. The outflows wrap around the compact disk. The right panel (c) shows the lab-frame radiation flux lines and radiation luminosity per unit angle. Most of the kinetic energy escapes through the funnel region, carried by strong outflows. The radiation flux leaked through this region can also be super-Eddington, as compared to being Eddington-limited at other inclination angles. }
\label{fig:energyflux}
\end{figure*}

\begin{figure*}
\figurenum{5}
\label{fig:spectra}
\plottwo{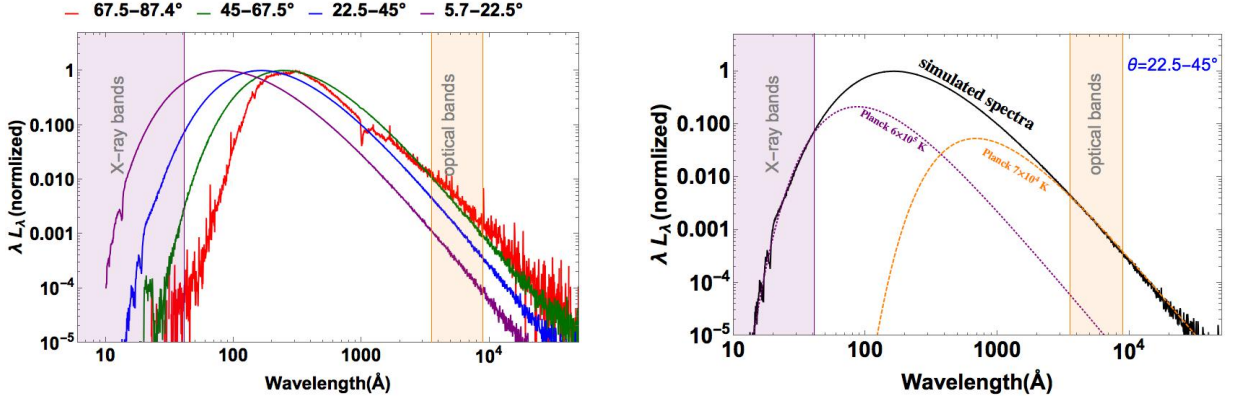}{CombineSpectra3.jpg}
\caption{(a) The simulated escaping spectra from inclination angles bin 1 (red curve, closest to mid-plane) to bin 4 (purple curve, closest to the pole). The optical to X-ray ratio increases with inclination angle. The purple shaded region corresponds to X-ray band with energy below 0.3 keV. The orange shaded region corresponds to the optical band from 3560\AA to 9000\AA. (b) The simulated escaping spectrum  (black curve) from intermediate angle bin 3, the shapes of which is broader than a blackbody spectrum. We also show two Planck functions, one fitting the X-ray continuum component (purple dotted curve) and the other one fitting the optical continuum component (orange dashed curve). 
}
\end{figure*}

The GRRMHD simulation was run up to a final time $t_f \sim 20000~R_g/c$, by which point the disk has achieved inflow equilibrium out to $r\sim~200~žR_g$ with constant fluxes of mass, energy, and specific angular momentum vs. radius. The disk $\alpha$-parameter is of order unity. Fig. \ref{fig:zoomin} shows a snapshot from the simulation as well as various fluxes and efficiencies vs. time. In the following, we focus on disk properties averaged over the latter quasi-steady stage of accretion, from $t = 15000 R_g /c$ until the end. The averaged accretion rate onto the black hole during this phase is $15~\dot{M}_{\rm Edd}$, close to the peak of the fallback rate of stellar debris $\dot{M}_{\rm fb}\sim 12~\dot{M}_{\rm Edd}$ when a solar type star is disrupted by a black hole of mass $5\times~10^{6}~M_\odot$. The total bolometric luminosity emitted is $L\sim~3.2~L_{\rm Edd}$, giving a radiative efficiency of $\eta_{\rm rad}~\sim~2.7\%$. The system is radiatively inefficient compared with the Novikov-Thorne thin disk solution. The total efficiency is $\eta_{\rm total}~\sim~43\%$. The jet efficiency $\eta_j~\sim~20\%$, and the efficiency of the outflow in terms of kinetic, thermal and gravitational binding energy is also $\sim~20\%$, both much larger than the radiative efficiency .

The initial disk is threaded with a weak poloidal magnetic field, which amplifies via the magneto-rotational instability. The magnetic flux accumulates and eventually the magnetically arrested disk (MAD) state is built up out to $r\sim~ž80~R_g$ during the latter quasi-steady stage. The magnetic flux accumulated in this region reaches $10^{31}~{\rm Guass~cm^2}$, which is consistent with the flux that the debris stream can capture from a fossil disk \citep{Kelley14}. Since the black hole is spinning fast, the strong magnetic field threading the BH and disk leads to a relativistic jet by extracting the black hole spin energy mainly through the Blandford-Znajek process \citep{Blandford77}.

A wide-angle, fast wind is launched from the magnetized disk supported by radiation pressure. Fig.~ \ref{fig:outflow} shows how the $t$-$\phi$-averaged gas density and radial velocity vary with the distance from the black hole along fixed inclination angles. The wind has drastically different density and velocity profiles at different inclination angles close to the black hole. At higher inclination angles (closer to mid plane), the outflows are denser, and travels at a velocity considerably smaller than the speed of light. At lower inclination angles, the outflows are much more dilute, and can have a radial velocity of few$\times0.1c$ (conventionally called ultra-fast outflows). The outflows are optically thick for the photons produced from the inner disk, so intrinsic emissions from the disk will be reprocessed in the outflows, as we will discuss in section \ref{sec:emission}. During the quasi-steady phase of the simulation, the outflows, mostly launched from the inner disk (where inflow equilibrium has been established), have travelled beyond the size of the electron scattering photosphere from most inclination angles. As shown in Fig. \ref{fig:energyflux}, the photosphere for electron scattering, where the optical depth $\tau_r~=~\int~\kappa_{\rm es}dl~=~1$ integrated from $r=8000~R_g$ (optically thin region) radially inwards, has been resolved. The photosphere has a size $\sim~1000~R_g$ from most viewing angles, but gets much closer to the black hole in the funnel region.

We show in Fig. \ref{fig:energyflux} the magnetic flux lines (with electromagnetic efficiency), the gas velocity flux lines (with kinetic+gravitational energy efficiency) and lab-frame radiation flux stream lines (with radiative efficiency). Within the optically thick region, photons are trapped in the gas by scattering, so radiation either moves inward within the disk or somewhat follows the path of the wind and ultimately becomes more radially-directed at larger distances. Most of the electromagnetic energy is released from the black hole's spin energy through the Blandford-Znajek process. Most of the electromagnetic and kinetic energy escape with the jet or through the wind in the funnel region. The radiation flux is mildly beamed -- the bolometric flux is around Eddington when viewed along the disk, and 2-3 times Eddington when looking down the funnel where strong outflows are present.

\subsection{Viewing angle dependence for emission}
\label{sec:emission}

Most of the radiation is produced from the inner disk and the base of the jet close to the black hole, which is reprocessed when going through the outflow and outer disk. Two different reprocessing mechanisms are responsible for changing the spectral energy distribution of the photons.  Away from the mid-plane, in inclination angle bins 2 through 4, the low density of the gas results in relatively high ionization state and therefore very little absorption of the radiation. Instead, it is trapped in the outflow by scattering, and its spectrum is red-shifted due to adiabatic expansion. Due to the high gas velocity, the thermal state of the electrons plays little role in setting the spectrum in these bins. 

Such adiabatic reprocessing also plays a role for gas close to the mid-plane (bin 1), but here the radiation is mainly absorbed by photoionization of HeII and OVI of the dense and slow gas in the outflows and disk. The absorbed radiation is re-emitted at UV and optical wavelengths. For the optical re-emission, the primary processes involved are recombination of free electrons onto hydrogen and free-free emission, along with some line emission, as in \citet{Roth16}.

We show the simulated escaping spectra from bin 1 to bin 4 in Fig. \ref{fig:spectra} (a). There is a clear trend that the optical to X-ray flux ratio increases with the inclination angle.  The ratios of the integrated optical/UV flux (1500--7000\AA) to X-ray flux (above 0.3 keV) are: 2700 (bin 1), 190 (bin 2), 2.1 (bin 3) and 0.036 (bin 4). In all bins, most of the escaping luminosity is in the EUV. The ratios of the (unobservable) EUV (41.3--1150\AA) to the combination of the (observable) optical and X-ray luminosities are: 24 (bin 1), 24 (bin 2), 39 (bin 3), and 7.1 (bin 4).  Interestingly, the shapes of the escaping spectra from all bins are broader than a blackbody spectrum.

\subsection{Comparison with TDE observations}
The simulated disk has various features consistent with the observed properties of TDEs. As our simulation represents only one epoch in the super-Eddington phase but could not follow the whole evolution of a TDE disk, we only compare with the properties observed around the peak of the TDE flare.

\hangindent=0.5cm
\hangafter=0
1. Fast outflow: Simulations show that the outflows from super-Eddington accretion can have ultra-fast speed of few$\times0.1c$ at relatively low inclination angles. Outflows with similarly high speed have been inferred in the jetted TDE \emph{Swift}-J1644 through X-ray reverberation \citep{Kara16}, as well as in the non-jetted (or weakly jetted) TDE ASASSN 14-li through X-ray absorption feature \citep{Kara17} and radio signals \citep{Alexander16}.

\hangindent=0.5cm
\hangafter=0
2. X-ray or optical TDEs: We cannot get a precise ratio of X-ray TDEs to optical TDEs, since parameters like black hole mass and spin and accretion rates in the observed TDEs could be different from the parameters used in this particular simulation. However, to the first order, we should expect that a large fraction of the optically-selected TDEs do not show strong thermal X-ray emissions, and vice versa. Only a small fraction of TDEs will show equally strong X-ray and optical emissions like ASASSN 14-li.

\hangindent=0.5cm
\hangafter=0
3. Disk temperature: The broad shape of the reprocessed spectra means that X-ray discovered TDEs will have a higher inferred blackbody temperature than optically discovered TDEs.  When both strong X-ray and optical emissions are observed from the same event, such as when viewed along bin 3, the X-ray part of the simulated spectra can be fitted by a Planck function of few $\times10^5$K, while the optical part can be fitted by a Planck function of few$\times10^4$K, as shown in Fig. \ref{fig:spectra} (b). This is consistent with what was observed in ASASSN 14-li and ASASSN 15-oi to the first order.

\hangindent=0.5cm
\hangafter=0
4. Luminosity: The luminosity estimate of several TDEs are a few$\times0.1 L_{\rm Edd}$ \citep{Hung17}. This is consistent with the luminosity inferred from the optical or X-ray component of the simulated spectra by fitting Planck functions to each of them separately, but underestimates the true bolometric luminosity. Given that most of the radiated energy is in EUV, which is likely to be absorbed by gas and dust in the host galaxy, perhaps the best hope of recovering an accurate estimate of the bolometric luminosity originally emitted by the TDE will be via observations of infrared echoes \citep[e.g.,][]{vanVelzen16b, Jiang16}.

\hangindent=0.5cm
\hangafter=0
5. Presence of jet: Only three TDEs have been observed to produce powerful relativistic jets. If these jets are produced through the Blandford-Znajek process, this can be because the host SMBHS of those events have very large spins, since the jet power generally depends on black hole spin as $P_j \propto a^{2-4}$ \citep{McKinney05}. It is also possible that only in those events were large magnetic fluxes accumulated or collected \citep{Kelley14}, given that the initial stellar magnetic flux is not enough to power a jet such as that from \emph{Swift}-J1644. The full answer to this question goes beyond the scope of this paper.

\section{Summary and future work}
\label{sec:summary}
We propose a unified model for various classes of observed TDEs (Fig. \ref{fig:basic}): The spectral distribution of emission observed from TDEs mainly depends on the the viewing angle of the observer with respect to the orientation of the disk. To examine this, we have carried out 3D GRRMHD simulations of a compact super-Eddington disk around a SMBH, with parameters consistent with realistic TDE scenarios. Wide-angle, optically thick outflows are launched from the radiation-pressure dominated thick accretion disk. Monte-Carlo radiative transfer studies using the simulation data reveal how the intrinsic emissions from inner disk are reprocessed in the outflows and the outer disk: 1) Close to the disk mid-plane, the intrinsic X-ray emission is absorbed by the dense, slow outflow or disk into NUV/optical emission by photoionization. 2) When viewed high above the disk, the dilute outflow moves at few$\times0.1$c, the gas is effectively optically thin to absorption, and adiabatic expansion is mainly responsible for cooling. 3) Only when looking down the funnel region would one expect to see the exposed inner disk producing strong, beamed X-ray emission. A relativistic jet can be produced when conditions are optimal, and the power of which could depend on the black hole spin and magnetic flux dragged into the accretion flow. Whether a jet is produced or not, however, does not change the basic picture that the the optical to X-ray flux ratio of the observed emission increases with the viewing angle of the observer.

It has been proposed that collisions of debris streams during the disk-formation phase can also produce optical emissions \citep{Piran15, Jiang16a}. Further studies are needed to compare the contribution of optical emissions from stream-stream collision and accretion. Line-driven winds might also contribute significantly to the outflow rate when the gas temperature is around $10^5$K \citep{MillerCole15} and should be considered. We focus on the continuum emission for this paper, and will investigate the line profiles with higher-resolution radiative transfer calculations in future work. Furthermore, simulations with different parameters are needed to understand the full evolution of a TDE. 

Much of what we have summarized in this study revolves around different ideas as to how three dimensional flows, likely magnetic and radiation dominated, behave in and around strong gravitational fields. While there are serious issues of theory that need to be settled, it is clear that there is a convergence in the study of AGN, radio jets, (ultra-luminous) X-ray binaries, and TDEs, which inspires the type of model as we advocated. These bonds may be highly relevant in teaching us how mass, angular momentum and energy can flow around and away from black holes.

\acknowledgments

We acknowledge support from the DNRF (L.D. and E.R ), NASA ATP grant NNX14AH37G and NSF grant AST-1615881 (E. R.). We thank Katie Auchettl, Erin Kara, and Brian Metzger as well as TCAN group members (Mark Avara, Roman Gold, Megan Marshall, Ramesh Narayan, Peter Polko, Aleksander S{\c a}dowski, Danilo Teixeira) for discussions. We also thank the anonymous referee for useful comments. LC and JCM acknowledge the computational support from NSF via XSEDE resources (TG-PHY120005 and TG-AST160003).


\begin{thebibliography}{}
\providecommand\natexlab[1]{#1}
\providecommand\JournalTitle[1]{#1}

\bibitem[{{Abramowicz} {et~al.}(1988){Abramowicz}, {Czerny}, {Lasota}, \&
  {Szuszkiewicz}}]{Abramowicz88}
{Abramowicz}, M.~A., {Czerny}, B., {Lasota}, J.~P., \& {Szuszkiewicz}, E. 1988,
  \href{http://dx.doi.org/10.1086/166683}{\JournalTitle{\apj}, 332, 646}

\bibitem[{{Alexander} {et~al.}(2016){Alexander}, {Berger}, {Guillochon},
  {Zauderer}, \& {Williams}}]{Alexander16}
{Alexander}, K.~D., {Berger}, E., {Guillochon}, J., {Zauderer}, B.~A., \&
  {Williams}, P.~K.~G. 2016,
  \href{http://dx.doi.org/10.3847/2041-8205/819/2/L25}{\JournalTitle{\apjl},
  819, L25}

\bibitem[{{Arcavi} {et~al.}(2014){Arcavi}, {Gal-Yam}, {Sullivan}, {Pan},
  {Cenko}, {Horesh}, {Ofek}, {De Cia}, {Yan}, {Yang}, {Howell}, {Tal},
  {Kulkarni}, {Tendulkar}, {Tang}, {Xu}, {Sternberg}, {Cohen}, {Bloom},
  {Nugent}, {Kasliwal}, {Perley}, {Quimby}, {Miller}, {Theissen}, \&
  {Laher}}]{Arcavi14}
{Arcavi}, I., {Gal-Yam}, A., {Sullivan}, M., {et~al.} 2014,
  \href{http://dx.doi.org/10.1088/0004-637X/793/1/38}{\JournalTitle{\apj}, 793,
  38}

\bibitem[{{Begelman}(1978)}]{Begelman78}
{Begelman}, M.~C. 1978,
  \href{http://dx.doi.org/10.1093/mnras/184.1.53}{\JournalTitle{\mnras}, 184,
  53}

\bibitem[{{Blandford} \& {Znajek}(1977)}]{Blandford77}
{Blandford}, R.~D., \& {Znajek}, R.~L. 1977,
  \href{http://dx.doi.org/10.1093/mnras/179.3.433}{\JournalTitle{\mnras}, 179,
  433}

\bibitem[{{Bloom} {et~al.}(2011){Bloom}, {Giannios}, {Metzger}, {Cenko},
  {Perley}, {Butler}, {Tanvir}, {Levan}, {O'Brien}, {Strubbe}, {De Colle},
  {Ramirez-Ruiz}, {Lee}, {Nayakshin}, {Quataert}, {King}, {Cucchiara},
  {Guillochon}, {Bower}, {Fruchter}, {Morgan}, \& {van der Horst}}]{Bloom11}
{Bloom}, J.~S., {Giannios}, D., {Metzger}, B.~D., {et~al.} 2011,
  \href{http://dx.doi.org/10.1126/science.1207150}{\JournalTitle{Science}, 333,
  203}

\bibitem[{{Bonnerot} {et~al.}(2016){Bonnerot}, {Rossi}, {Lodato}, \&
  {Price}}]{Bonnerot16}
{Bonnerot}, C., {Rossi}, E.~M., {Lodato}, G., \& {Price}, D.~J. 2016,
  \href{http://dx.doi.org/10.1093/mnras/stv2411}{\JournalTitle{\mnras}, 455,
  2253}

\bibitem[{{Brown} {et~al.}(2015){Brown}, {Levan}, {Stanway}, {Tanvir}, {Cenko},
  {Berger}, {Chornock}, \& {Cucchiaria}}]{Brown15}
{Brown}, G.~C., {Levan}, A.~J., {Stanway}, E.~R., {et~al.} 2015,
  \href{http://dx.doi.org/10.1093/mnras/stv1520}{\JournalTitle{\mnras}, 452,
  4297}

\bibitem[{{Burrows} {et~al.}(2011){Burrows}, {Kennea}, {Ghisellini}, {Mangano},
  {Zhang}, {Page}, {Eracleous}, {Romano}, {Sakamoto}, {Falcone}, {Osborne},
  {Campana}, {Beardmore}, {Breeveld}, {Chester}, {Corbet}, {Covino},
  {Cummings}, {D'Avanzo}, {D'Elia}, {Esposito}, {Evans}, {Fugazza}, {Gelbord},
  {Hiroi}, {Holland}, {Huang}, {Im}, {Israel}, {Jeon}, {Jeon}, {Jun}, {Kawai},
  {Kim}, {Krimm}, {Marshall}, {P.~M{\'e}sz{\'a}ros}, {Negoro}, {Omodei},
  {Park}, {Perkins}, {Sugizaki}, {Sung}, {Tagliaferri}, {Troja}, {Ueda},
  {Urata}, {Usui}, {Antonelli}, {Barthelmy}, {Cusumano}, {Giommi}, {Melandri},
  {Perri}, {Racusin}, {Sbarufatti}, {Siegel}, \& {Gehrels}}]{Burrows11}
{Burrows}, D.~N., {Kennea}, J.~A., {Ghisellini}, G., {et~al.} 2011,
  \href{http://dx.doi.org/10.1038/nature10374}{\JournalTitle{\nat}, 476, 421}

\bibitem[{{Cannizzo} {et~al.}(1990){Cannizzo}, {Lee}, \&
  {Goodman}}]{Cannizzo90}
{Cannizzo}, J.~K., {Lee}, H.~M., \& {Goodman}, J. 1990,
  \href{http://dx.doi.org/10.1086/168442}{\JournalTitle{\apj}, 351, 38}

\bibitem[{{Cenko} {et~al.}(2012){Cenko}, {Krimm}, {Horesh}, {Rau}, {Frail},
  {Kennea}, {Levan}, {Holland}, {Butler}, {Quimby}, {Bloom}, {Filippenko},
  {Gal-Yam}, {Greiner}, {Kulkarni}, {Ofek}, {Olivares E.}, {Schady},
  {Silverman}, {Tanvir}, \& {Xu}}]{Cenko12}
{Cenko}, S.~B., {Krimm}, H.~A., {Horesh}, A., {et~al.} 2012,
  \href{http://dx.doi.org/10.1088/0004-637X/753/1/77}{\JournalTitle{\apj}, 753,
  77}

\bibitem[{{Coughlin} \& {Begelman}(2014)}]{Coughlin14}
{Coughlin}, E.~R., \& {Begelman}, M.~C. 2014,
  \href{http://dx.doi.org/10.1088/0004-637X/781/2/82}{\JournalTitle{\apj}, 781,
  82}

\bibitem[{{Dai} {et~al.}(2013){Dai}, {Escala}, \& {Coppi}}]{Dai13}
{Dai}, L., {Escala}, A., \& {Coppi}, P. 2013,
  \href{http://dx.doi.org/10.1088/2041-8205/775/1/L9}{\JournalTitle{\apjl},
  775, L9}

\bibitem[{{Dai} {et~al.}(2015){Dai}, {McKinney}, \& {Miller}}]{Dai15}
{Dai}, L., {McKinney}, J.~C., \& {Miller}, M.~C. 2015,
  \href{http://dx.doi.org/10.1088/2041-8205/812/2/L39}{\JournalTitle{\apjl},
  812, L39}

\bibitem[{{Evans} \& {Kochanek}(1989)}]{Evans89}
{Evans}, C.~R., \& {Kochanek}, C.~S. 1989,
  \href{http://dx.doi.org/10.1086/185567}{\JournalTitle{\apjl}, 346, L13}

\bibitem[{{Gezari} {et~al.}(2012){Gezari}, {Chornock}, {Rest}, {Huber},
  {Forster}, {Berger}, {Challis}, {Neill}, {Martin}, {Heckman}, {Lawrence},
  {Norman}, {Narayan}, {Foley}, {Marion}, {Scolnic}, {Chomiuk}, {Soderberg},
  {Smith}, {Kirshner}, {Riess}, {Smartt}, {Stubbs}, {Tonry}, {Wood-Vasey},
  {Burgett}, {Chambers}, {Grav}, {Heasley}, {Kaiser}, {Kudritzki}, {Magnier},
  {Morgan}, \& {Price}}]{Gezari12}
{Gezari}, S., {Chornock}, R., {Rest}, A., {et~al.} 2012,
  \href{http://dx.doi.org/10.1038/nature10990}{\JournalTitle{\nat}, 485, 217}

\bibitem[{{Guillochon} \& {Ramirez-Ruiz}(2013)}]{Guillochon13}
{Guillochon}, J., \& {Ramirez-Ruiz}, E. 2013,
  \href{http://dx.doi.org/10.1088/0004-637X/767/1/25}{\JournalTitle{\apj}, 767,
  25}

\bibitem[{{Guillochon} \& {Ramirez-Ruiz}(2015)}]{Guillochon15}
---. 2015,
  \href{http://dx.doi.org/10.1088/0004-637X/809/2/166}{\JournalTitle{\apj},
  809, 166}

\bibitem[{{Hayasaki} {et~al.}(2016){Hayasaki}, {Stone}, \& {Loeb}}]{Hayasaki16}
{Hayasaki}, K., {Stone}, N., \& {Loeb}, A. 2016,
  \href{http://dx.doi.org/10.1093/mnras/stw1387}{\JournalTitle{\mnras}, 461,
  3760}

\bibitem[{{Hills}(1975)}]{Hills75}
{Hills}, J.~G. 1975,
  \href{http://dx.doi.org/10.1038/254295a0}{\JournalTitle{\nat}, 254, 295}

\bibitem[{{Holoien} {et~al.}(2016{\natexlab{a}}){Holoien}, {Kochanek},
  {Prieto}, {Grupe}, {Chen}, {Godoy-Rivera}, {Stanek}, {Shappee}, {Dong},
  {Brown}, {Basu}, {Beacom}, {Bersier}, {Brimacombe}, {Carlson}, {Falco},
  {Johnston}, {Madore}, {Pojmanski}, \& {Seibert}}]{Holoien16_15oi}
{Holoien}, T.~W.-S., {Kochanek}, C.~S., {Prieto}, J.~L., {et~al.}
  2016{\natexlab{a}},
  \href{http://dx.doi.org/10.1093/mnras/stw2272}{\JournalTitle{\mnras}, 463,
  3813}

\bibitem[{{Holoien} {et~al.}(2016{\natexlab{b}}){Holoien}, {Kochanek},
  {Prieto}, {Stanek}, {Dong}, {Shappee}, {Grupe}, {Brown}, {Basu}, {Beacom},
  {Bersier}, {Brimacombe}, {Danilet}, {Falco}, {Guo}, {Jose}, {Herczeg},
  {Long}, {Pojmanski}, {Simonian}, {Szczygie{\l}}, {Thompson}, {Thorstensen},
  {Wagner}, \& {Wo{\'z}niak}}]{Holoien16_14li}
---. 2016{\natexlab{b}},
  \href{http://dx.doi.org/10.1093/mnras/stv2486}{\JournalTitle{\mnras}, 455,
  2918}

\bibitem[{{Hung} {et~al.}(2017){Hung}, {Gezari}, {Blagorodnova}, {Roth},
  {Cenko}, {Kulkarni}, {Horesh}, {Arcavi}, {McCully}, {Yan}, {Lunnan},
  {Fremling}, {Cao}, {Nugent}, \& {Wozniak}}]{Hung17}
{Hung}, T., {Gezari}, S., {Blagorodnova}, N., {et~al.} 2017,
  \href{http://dx.doi.org/10.3847/1538-4357/aa7337}{\JournalTitle{\apj}, 842,
  29}

\bibitem[{{Jiang} {et~al.}(2016{\natexlab{a}}){Jiang}, {Dou}, {Wang}, {Yang},
  {Lyu}, \& {Zhou}}]{Jiang16}
{Jiang}, N., {Dou}, L., {Wang}, T., {et~al.} 2016{\natexlab{a}},
  \href{http://dx.doi.org/10.3847/2041-8205/828/1/L14}{\JournalTitle{\apjl},
  828, L14}

\bibitem[{{Jiang} {et~al.}(2016{\natexlab{b}}){Jiang}, {Guillochon}, \&
  {Loeb}}]{Jiang16a}
{Jiang}, Y.-F., {Guillochon}, J., \& {Loeb}, A. 2016{\natexlab{b}},
  \href{http://dx.doi.org/10.3847/0004-637X/830/2/125}{\JournalTitle{\apj},
  830, 125}

\bibitem[{{Jiang} {et~al.}(2014){Jiang}, {Stone}, \& {Davis}}]{Jiang14}
{Jiang}, Y.-F., {Stone}, J.~M., \& {Davis}, S.~W. 2014,
  \href{http://dx.doi.org/10.1088/0004-637X/796/2/106}{\JournalTitle{\apj},
  796, 106}

\bibitem[{{Kara} {et~al.}(2018){Kara}, {Dai}, {Reynolds}, \&
  {Kallman}}]{Kara17}
{Kara}, E., {Dai}, L., {Reynolds}, C.~S., \& {Kallman}, T. 2018,
  \href{http://dx.doi.org/10.1093/mnras/stx3004}{\JournalTitle{\mnras}, 474,
  3593}

\bibitem[{{Kara} {et~al.}(2016){Kara}, {Miller}, {Reynolds}, \& {Dai}}]{Kara16}
{Kara}, E., {Miller}, J.~M., {Reynolds}, C., \& {Dai}, L. 2016,
  \href{http://dx.doi.org/10.1038/nature18007}{\JournalTitle{\nat}, 535, 388}

\bibitem[{{Kasen} {et~al.}(2006){Kasen}, {Thomas}, \& {Nugent}}]{Kasen06}
{Kasen}, D., {Thomas}, R.~C., \& {Nugent}, P. 2006,
  \href{http://dx.doi.org/10.1086/506190}{\JournalTitle{\apj}, 651, 366}

\bibitem[{{Kelley} {et~al.}(2014){Kelley}, {Tchekhovskoy}, \&
  {Narayan}}]{Kelley14}
{Kelley}, L.~Z., {Tchekhovskoy}, A., \& {Narayan}, R. 2014,
  \href{http://dx.doi.org/10.1093/mnras/stu2041}{\JournalTitle{\mnras}, 445,
  3919}

\bibitem[{{Komossa}(2015)}]{Komossa15review}
{Komossa}, S. 2015,
  \href{http://dx.doi.org/10.1016/j.jheap.2015.04.006}{\JournalTitle{Journal of
  High Energy Astrophysics}, 7, 148}

\bibitem[{{Levan} {et~al.}(2011){Levan}, {Tanvir}, {Cenko}, {Perley},
  {Wiersema}, {Bloom}, {Fruchter}, {Postigo}, {O'Brien}, {Butler}, {van der
  Horst}, {Leloudas}, {Morgan}, {Misra}, {Bower}, {Farihi}, {Tunnicliffe},
  {Modjaz}, {Silverman}, {Hjorth}, {Th{\"o}ne}, {Cucchiara}, {Cer{\'o}n},
  {Castro-Tirado}, {Arnold}, {Bremer}, {Brodie}, {Carroll}, {Cooper}, {Curran},
  {Cutri}, {Ehle}, {Forbes}, {Fynbo}, {Gorosabel}, {Graham}, {Hoffman},
  {Guziy}, {Jakobsson}, {Kamble}, {Kerr}, {Kasliwal}, {Kouveliotou},
  {Kocevski}, {Law}, {Nugent}, {Ofek}, {Poznanski}, {Quimby}, {Rol},
  {Romanowsky}, {S{\'a}nchez-Ram{\'{\i}}rez}, {Schulze}, {Singh}, {van
  Spaandonk}, {Starling}, {Strom}, {Tello}, {Vaduvescu}, {Wheatley}, {Wijers},
  {Winters}, \& {Xu}}]{Levan11}
{Levan}, A.~J., {Tanvir}, N.~R., {Cenko}, S.~B., {et~al.} 2011,
  \href{http://dx.doi.org/10.1126/science.1207143}{\JournalTitle{Science}, 333,
  199}

\bibitem[{{Levermore}(1984)}]{Levermore84}
{Levermore}, C.~D. 1984,
  \href{http://dx.doi.org/10.1016/0022-4073(84)90112-2}{\JournalTitle{\jqsrt},
  31, 149}

\bibitem[{{Lodato} {et~al.}(2009){Lodato}, {King}, \& {Pringle}}]{Lodato09}
{Lodato}, G., {King}, A.~R., \& {Pringle}, J.~E. 2009,
  \href{http://dx.doi.org/10.1111/j.1365-2966.2008.14049.x}{\JournalTitle{\mnras},
  392, 332}

\bibitem[{{Loeb} \& {Ulmer}(1997)}]{Loeb97}
{Loeb}, A., \& {Ulmer}, A. 1997,
  \href{http://dx.doi.org/10.1086/304814}{\JournalTitle{\apj}, 489, 573}

\bibitem[{{McKinney}(2005)}]{McKinney05}
{McKinney}, J.~C. 2005,
  \href{http://dx.doi.org/10.1086/468184}{\JournalTitle{\apjl}, 630, L5}

\bibitem[{{McKinney} {et~al.}(2015){McKinney}, {Dai}, \& {Avara}}]{McKinney15}
{McKinney}, J.~C., {Dai}, L., \& {Avara}, M.~J. 2015,
  \href{http://dx.doi.org/10.1093/mnrasl/slv115}{\JournalTitle{\mnras}, 454,
  L6}

\bibitem[{{McKinney} {et~al.}(2014){McKinney}, {Tchekhovskoy}, {Sadowski}, \&
  {Narayan}}]{McKinney14}
{McKinney}, J.~C., {Tchekhovskoy}, A., {Sadowski}, A., \& {Narayan}, R. 2014,
  \href{http://dx.doi.org/10.1093/mnras/stu762}{\JournalTitle{\mnras}, 441,
  3177}

\bibitem[{{Metzger} \& {Stone}(2016)}]{Metzger16}
{Metzger}, B.~D., \& {Stone}, N.~C. 2016,
  \href{http://dx.doi.org/10.1093/mnras/stw1394}{\JournalTitle{\mnras}, 461,
  948}

\bibitem[{{Miller} {et~al.}(2015){Miller}, {Kaastra}, {Miller}, {Reynolds},
  {Brown}, {Cenko}, {Drake}, {Gezari}, {Guillochon}, {Gultekin}, {Irwin},
  {Levan}, {Maitra}, {Maksym}, {Mushotzky}, {O'Brien}, {Paerels}, {de Plaa},
  {Ramirez-Ruiz}, {Strohmayer}, \& {Tanvir}}]{Miller15}
{Miller}, J.~M., {Kaastra}, J.~S., {Miller}, M.~C., {et~al.} 2015,
  \href{http://dx.doi.org/10.1038/nature15708}{\JournalTitle{\nat}, 526, 542}

\bibitem[{{Miller}(2015)}]{MillerCole15}
{Miller}, M.~C. 2015,
  \href{http://dx.doi.org/10.1088/0004-637X/805/1/83}{\JournalTitle{\apj}, 805,
  83}

\bibitem[{{Mockler} {et~al.}(2018){Mockler}, {Guillochon}, \&
  {Ramirez-Ruiz}}]{Mockler18}
{Mockler}, B., {Guillochon}, J., \& {Ramirez-Ruiz}, E. 2018,
  \JournalTitle{ArXiv e-prints},
  \href{http://arxiv.org/abs/1801.08221}{{\sffamily arXiv:1801.08221
  [astro-ph.HE]}}

\bibitem[{{Novikov} \& {Thorne}(1973)}]{Novikov73}
{Novikov}, I.~D., \& {Thorne}, K.~S. 1973, in Black Holes (Les Astres Occlus),
  ed. C.~{Dewitt} \& B.~S. {Dewitt}, 343

\bibitem[{{Ohsuga} {et~al.}(2005){Ohsuga}, {Mori}, {Nakamoto}, \&
  {Mineshige}}]{Ohsuga05}
{Ohsuga}, K., {Mori}, M., {Nakamoto}, T., \& {Mineshige}, S. 2005,
  \href{http://dx.doi.org/10.1086/430728}{\JournalTitle{\apj}, 628, 368}

\bibitem[{{Phinney}(1989)}]{Phinney89}
{Phinney}, E.~S. 1989, in IAU Symposium, Vol. 136, The Center of the Galaxy,
  ed. M.~{Morris}, 543

\bibitem[{{Piran} {et~al.}(2015){Piran}, {Svirski}, {Krolik}, {Cheng}, \&
  {Shiokawa}}]{Piran15}
{Piran}, T., {Svirski}, G., {Krolik}, J., {Cheng}, R.~M., \& {Shiokawa}, H.
  2015,
  \href{http://dx.doi.org/10.1088/0004-637X/806/2/164}{\JournalTitle{\apj},
  806, 164}

\bibitem[{{Rees}(1988)}]{Rees88}
{Rees}, M.~J. 1988,
  \href{http://dx.doi.org/10.1038/333523a0}{\JournalTitle{\nat}, 333, 523}

\bibitem[{{Roth} \& {Kasen}(2017)}]{Roth17}
{Roth}, N., \& {Kasen}, D. 2017, \JournalTitle{ArXiv e-prints},
  \href{http://arxiv.org/abs/1707.02993}{{\sffamily arXiv:1707.02993
  [astro-ph.HE]}}

\bibitem[{{Roth} {et~al.}(2016){Roth}, {Kasen}, {Guillochon}, \&
  {Ramirez-Ruiz}}]{Roth16}
{Roth}, N., {Kasen}, D., {Guillochon}, J., \& {Ramirez-Ruiz}, E. 2016,
  \href{http://dx.doi.org/10.3847/0004-637X/827/1/3}{\JournalTitle{\apj}, 827,
  3}

\bibitem[{{S{\c a}dowski} {et~al.}(2015){S{\c a}dowski}, {Narayan},
  {Tchekhovskoy}, {Abarca}, {Zhu}, \& {McKinney}}]{Sadowski15}
{S{\c a}dowski}, A., {Narayan}, R., {Tchekhovskoy}, A., {et~al.} 2015,
  \href{http://dx.doi.org/10.1093/mnras/stu2387}{\JournalTitle{\mnras}, 447,
  49}

\bibitem[{{Shiokawa} {et~al.}(2015){Shiokawa}, {Krolik}, {Cheng}, {Piran}, \&
  {Noble}}]{Shiokawa15}
{Shiokawa}, H., {Krolik}, J.~H., {Cheng}, R.~M., {Piran}, T., \& {Noble}, S.~C.
  2015,
  \href{http://dx.doi.org/10.1088/0004-637X/804/2/85}{\JournalTitle{\apj}, 804,
  85}

\bibitem[{{Strubbe} \& {Quataert}(2009)}]{Strubbe09}
{Strubbe}, L.~E., \& {Quataert}, E. 2009,
  \href{http://dx.doi.org/10.1111/j.1365-2966.2009.15599.x}{\JournalTitle{\mnras},
  400, 2070}

\bibitem[{{Tchekhovskoy} {et~al.}(2014){Tchekhovskoy}, {Metzger}, {Giannios},
  \& {Kelley}}]{Tchekhovskoy14}
{Tchekhovskoy}, A., {Metzger}, B.~D., {Giannios}, D., \& {Kelley}, L.~Z. 2014,
  \href{http://dx.doi.org/10.1093/mnras/stt2085}{\JournalTitle{\mnras}, 437,
  2744}

\bibitem[{{Ulmer}(1999)}]{Ulmer99}
{Ulmer}, A. 1999, \href{http://dx.doi.org/10.1086/306909}{\JournalTitle{\apj},
  514, 180}

\bibitem[{{van Velzen} {et~al.}(2016){van Velzen}, {Mendez}, {Krolik}, \&
  {Gorjian}}]{vanVelzen16b}
{van Velzen}, S., {Mendez}, A.~J., {Krolik}, J.~H., \& {Gorjian}, V. 2016,
  \href{http://dx.doi.org/10.3847/0004-637X/829/1/19}{\JournalTitle{\apj}, 829,
  19}

\end{thebibliography}
\end{document}